\documentclass[12pt]{article} % 10pt is ignored!

\usepackage{epsfig,bbm}

\renewcommand{\thefootnote}{\fnsymbol{footnote}}  
\usepackage{ifthen}
\usepackage{pifont}
\usepackage{epsfig}
\usepackage{amssymb}
\usepackage{graphicx}
\usepackage{amsmath}
\usepackage[center]{subfigure}
\newcommand{\ba}{\begin{eqnarray}}
\newcommand{\ea}{\end{eqnarray}}
\newcommand{\be}{\begin{equation}}
\newcommand{\ee}{\end{equation}}

\makeatletter
\def\fmslash{\@ifnextchar[{\fmsl@sh}{\fmsl@sh[0mu]}}
\def\fmsl@sh[#1]#2{%
  \mathchoice
    {\@fmsl@sh\displaystyle{#1}{#2}}%
    {\@fmsl@sh\textstyle{#1}{#2}}%
    {\@fmsl@sh\scriptstyle{#1}{#2}}%
    {\@fmsl@sh\scriptscriptstyle{#1}{#2}}}
\def\@fmsl@sh#1#2#3{\m@th\ooalign{$\hfil#1\mkern#2/\hfil$\crcr$#1#3$}}
\makeatother
\begin{document}

%%%%%%%%%% Title page   
\begin{titlepage}   
%\begin{flushright}   
%\end{flushright}   
\vskip1.5cm   

\begin{center}   

{\Large \bf \boldmath   
%CP VIOLATION AND THE 4th GENERATION}   
CP violation and the 4th generation}
\vskip2.0cm    
{\sc   
Gad Eilam    \footnote{eilam@physics.technion.ac.il}$^{,1}$,   
Bla\v zenka Meli\'c      \footnote{melic@thphys.irb.hr}$^{,2}$,   
\\ and \\   
Josip Trampeti\'c        \footnote{josipt@rex.irb.hr}$^{,3}$   
}
\vskip1.0cm   
$^1$  Department of Physics, Technion, Israel Institute of Technology, Haifa, 32000, Israel
\\
$^2$  Rudjer Bo\v skovi\' c Institute, Theoretical Physics Division, Bijeni\v cka 54,
\\HR-10002 Zagreb, Croatia
\\ 
$^3$  Rudjer Bo\v skovi\' c Institute, Bijeni\v cka 54,
HR-10002 Zagreb, Croatia
   
\vskip3cm  
 
%{\em Version of \today}   
   
\vskip1cm   
   
{\large\bf Abstract\\[10pt]} \parbox[t]{\textwidth}
{Within the Standard model with the 4th generation quarks $b'$ and $t'$ 
we have analyzed CP-violating flavor changing neutral current processes
$t \to c\,X,\;b^{\prime} \to s\,X,\;b^{\prime} \to b\,X,\;t^{\prime} \to c\,X$,
and $t^{\prime} \to t\,X$, with $X=H,Z,\gamma,g$, by constructing and employing global,
unique fit for the 4th generation mass mixing matrix CKM4 at $300 \leq m_{t'} \leq 700$ GeV.
All quantities appearing in the CKM4 were subject to our fitting procedure.
We have found that our fit produces the following CP partial rate asymmetry dominance: 
$a_{CP}(b^{\prime} \to s\,(H,Z;\gamma,g))\simeq\,(94,62;47,41)$ \%, at
$m_{t'}\simeq 300;350$ GeV, respectively.
From the experimental point of view the best decay mode,
out of the above four, is certainly $b^{\prime} \to s\,\gamma$, 
due to the presence of the high energy
single photon in the final state. We have also obtained relatively large asymmetry 
$a_{CP}(t \to cg) \simeq (8-18) $ \% for $t'$ running in the loops. 
There are fair chances that the 4th generation quarks
will be discovered at LHC and that some of their decay rates shall be measured.
If $b'$ and $t'$ exist at energies we assumed, with well executed tagging, large $a_{CP}$
could be found too.}   
   
\vfill   
   
\end{center}   
\end{titlepage}   
   
\setcounter{footnote}{0}   
\renewcommand{\thefootnote}{\arabic{footnote}}   
\renewcommand{\theequation}{\arabic{section}.\arabic{equation}}   
   
%\pacs{11.30.Er, 11.30.Hv, 12.60.-i}
%\keywords{fourth generation, CP, beyond SM \\ \vspace{3cm} \hspace{24cm} \it{\today}}

\newpage   
   
%%%%%%%%%%%%%%%%%%%%%%%%%%%%%%%%%%%%%%% 

\section{Introduction}   
\setcounter{equation}{0}

In this paper the main idea is to find possibly large genuine CP violation (CPV) effects
in the decays of the fourth generation of quarks,
arising from the one-loop flavor changing neutral currents (FCNC), 
by using unique fitting procedure.

A fourth generation of quarks and leptons, which we refer as 
$(t^{\prime},b^{\prime},\ell^{\prime},\nu^{\prime})$,
in our opinion is one of the most 
conservative guesses one could make as to what new physics lies ahead. Since
the 4th generation of flavors is neither 
predicted nor disallowed by the Standard Model (SM3)
we should keep an open mind regarding its existence.

Since possible existence of the 4th
generation provides a number of desirable features, 
for a review see \cite{Frampton:1999xi} and
for a more updated review see \cite{Holdom:2009rf}, 
let us first remind the reader of those.

$\star$ Fourth family is consistent with electroweak (EW) precision tests
\cite{Novikov:2001md}, because, if the 4th generation of fermions satisfy 
the following constraint for quarks \cite{Kribs:2007nz}
\begin{equation}
m_{t'}-m_{b'}\simeq\left(1+\frac{1}{5}\ln \frac{m_H}{115\,\rm GeV}\right)\times55\;\;\rm GeV\,,
\label{EWmass1}
\end{equation}
and the related 4th lepton generation mass difference 
%\begin{eqnarray*}
$m_{l'} - m_{\nu'} \simeq 60 \; {\rm GeV}$ \cite{Kribs:2007nz},  
%\end{eqnarray*}
the electroweak oblique parameters \cite{Peskin:1991sw} are not extending the experimentally
allowed parameter space.

$\star$ The 4th family of fermions is consistent with SU(5) gauge coupling, i.e.
it can be unified without supersymmetry, and, because of (\ref{EWmass1}), 
the 4th generation softens the current Higgs bounds \cite{Hung:1997zj}.

$\star$ With respect to quark-neutrino physics, if the 4th generation were discovered, it 
may change our prediction for the
$K^+ \to \pi^+ \nu\bar\nu$ decay \cite{Hou:1986ug}. 
There can be also implications on other penguin-induced decays, like 
$b \to s\gamma$ and $b \to s \phi$,  
see \cite{Deshpande:1989vk} and \cite{Deshpande:1989kc}, respectively.

$\star$ A heavy 4th family could naturally play a role in the dynamical breaking of EW symmetry
\cite{Hou:1987vd,Holdom:1986rn}. 

$\star$ If the unitarity of SM3 CKM matrix $V^{}_{3 \times 3}=V_{\rm CKM3}$ is slightly broken, 
new informations from top-quark production
at Tevatron still leaves open possibility that $|V_{tb}|$ is nontrivialy smaller
than one \cite{Alwall:2006bx}.

$\star$ In addition, a new generation might also cure some flavor physics problems too
\cite{Hou:1987vd,Hou:2005hd}.

$\star$  Finally, the 4th family might solve baryogenesis related problems, 
by visible increase of the
measure of CP violation and the strength of the phase transition.
Namely, the question of the large CP-violation in the SM3 extended to the 4th
generation of fermions (SM4) was recently raised in \cite{Hou:2008xd},
with respect of the insufficient
CP asymmetry produced in the Standard model with three generation 
for generating the baryon asymmetry in the universe.

There is a known quantity, the Jarlskog invariant, which measures the CP violation in the model.
For SM3, it is defined as \cite{Jarlskog:1985ht}
\begin{equation}
J = (m_t^2 - m_u^2)(m_t^2 - m_c^2)(m_c^2 - m_u^2)(m_b^2 - m_d^2)(m_b^2 - m_s^2)(m_s^2 - m_d^2) A\,,
\label{1}
\end{equation}
where $A$ is the twice of the area of any of six unitary triangles in SM3 and it is of order
of $O(10^{-5})$.

The fourth generation is added to the three known $SU(2)_f$ doublets as
\begin{equation}
\left ( \begin{array}{c} t' \\
                      b'
\end{array}
\right )\,.
\label{2}
\end{equation}
So in the case of
SM4, the above relation (\ref{1}) in the $d-s$ degeneracy limit generalizes to \cite{Hou:2008xd}
\begin{equation}
J_{234}^{bs} = (m_{t'}^2 - m_c^2)(m_{t'}^2 - m_t^2)(m_t^2 - m_c^2)
(m_{b'}^2 - m_s^2)(m_{b'}^2 - m_b^2)(m_b^2 - m_s^2) A_{234}^{bs}\,,
\label{Jarlskog}
\end{equation}
which, due to the large $m_{b'}, m_{t'}$ masses and somewhat larger area of the $b \to s$
quadrangle $A_{234}^{bs}$ corresponding to the SM4 unitarity relation
$V^{}_{4 \times 4} V^{\dagger}_{4 \times 4} = 1$,
can be up to 15 orders of magnitude larger than the Jarlskog invariant in the SM3 \cite{Hou:2008xd}.
However, this enhancement of CPV in the model with 4th fermion generation
cannot, by itself, solve the problem of the baryogenesis, since just adding the fermions
reduces the electroweak phase transition if there are not some additional theories involved like supersymmetry 
\cite{Fok:2008yg}, or the theory with at least two Higgs doublets \cite{Kikukawa:2009mu}. 

We start our analysis by performing a fit of SM4 CKM mass matrix. 
To obtain valid CPV results it is also crucial to follow results of the general
fit of the electroweak precision data, since, in order to be able to calculate
the real value of $A_{234}^{bs}$, 
it is necessary to make a global fit of a complete 
$V_{4 \times 4}$ matrix. 
Therefore, on top of many different processes used in the fit, we have also taken 
into account the EW constraints on the CKM mixing between the 3rd and the 4th quark family. 

We than compute flavor 
changing neutral current (FCNC) processes of the fourth generation quarks,
like $b' \to s (H,Z,\gamma,g)$, etc., and
analyze the most important consequence: large CP-violation in such decays.
The rare top decays $t \to c (H,Z,\gamma,g)$, 
involving the 4th generation quarks running in the loops are considered too. 

The paper is organized as follows. In Section 2. we introduce the quark mixing matrix
with 4th generation and construct the fitting procedure,
the fit itself, and present the corresponding results, respectively.
Section 3. contains computations of the FCNC processes
involving 4th family, while the CP violation effects due to the 4th generation are discussed
in Section 4. Lastly, Section 5. is devoted to discussions and a conclusion.

%%%%%%%%%%%%%%%%%%%%%%%%%%%%%%%%%%%%%%%%%%%%%%%%%%%%%%%%%%%%%
\section{CKM matrix for the fourth generation}
\setcounter{equation}{0}   

The fourth generation $4 \times 4$ quark mixing matrix is given by 
\begin{equation}
V_{4\times 4}\equiv V_{\rm CKM4} = \left (
\begin{array}{cccc} 
V_{ud} & V_{us} & V_{ub} & V_{ub'} \\
V_{cd} & V_{cs} & V_{cb} & V_{cb'} \\
V_{td} & V_{ts} & V_{tb} & V_{tb'} \\
V_{t'd} & V_{t's} & V_{t'b} & V_{t'b'}
\end{array}
\right )\,.
\label{4}
\end{equation}
The parameterization of such a matrix can be done in many possible ways. We have chosen to 
use the standard CKM3 Wolfenstein parameterization \cite{Wolfenstein:1983yz} 
of the $3 \times 3$ matrix, up to $O(\lambda^5)$%
\begin{eqnarray}
V_{\rm CKM3} &=& \left (
\begin{array}{ccc}
V_{ud} & V_{us} & V_{ub}  \\
V_{cd} & V_{cs} & V_{cb}  \\
V_{td} & V_{ts} & V_{tb}  
\end{array}
\right ) 
\nonumber \\
&& \hspace*{-1.8cm}  = 
\left( 
\begin{array}{ccc}
1 - \frac{\lambda^2}{2} - \frac{\lambda^4}{8} & \lambda & A \lambda^3 (\rho - i \eta) 
\\
\lambda\left(-1+A^2\frac{\lambda^4}{2}\left(1-2(\rho+i\eta)\right)\right) 
& 1-\frac{\lambda^2}{2}-\frac{\lambda^4}{8}(1+4A^2) 
& A \lambda^2 
\\
A\lambda^3 \left ( 1 - (\rho + i \eta) \left ( 1- \frac{\lambda^2}{2} \right ) \right ) &
 A \lambda^2 \left(-1 + \frac{\lambda^2}{2} \left( 1- 2( \rho + i \eta )\right)\right)  &
1 - A^2 \frac{\lambda^4}{2}
\end{array}
\right),
\label{CKM3}
\end{eqnarray}
and to multiply it by the 
mixing matrices of the first, the second and the third generation with the fourth generation,
 $R_{14}, R_{24}, R_{34}$, respectively, in a following way \cite{Botella:1985gb}:
\begin{equation}
V_{\rm CKM4} =  R_{34} \cdot R_{24} \cdot R_{14} \cdot V_{\rm CKM3}\,,
\label{CKM4def}
\end{equation}
where
\begin{eqnarray}
R_{34} &=& \left (
\begin{array}{cccc}
1 & 0 & 0 & 0 \\
0 & 1 & 0 & 0 \\
0 & 0 & c_u & s_u \\
0 & 0 & -s_u & c_u \\
\end{array}
\right),
\label{34}\\
R_{24} &=& \left (
\begin{array}{cccc}
1 & 0 & 0 & 0 \\
0 & c_v  & 0 & s_v e^{- i \phi_2} \\
0 & 0 & 1 & 0 \\
0 & -s_v e^{i \phi_2} & 0 &  c_v 
\end{array}
\right),
\label{24}\\
R_{14} &=& \left (
\begin{array}{cccc}
c_w & 0  & 0 & s_w e^{- i \phi_3} \\
0 & 1 & 0 & 0 \\
0 & 0 & 1 & 0 \\
-s_w e^{i \phi_3} & 0 & 0 & c_w 
\end{array}
\right).
\label{14}
\end{eqnarray}
The values for the parameters $\lambda$, $A$, $\rho$ and $\eta$ are taken to be in 
the range given by the 
global fit in SM3, \cite{Amsler:2008zzb}. 
The fourth generation parameters, $c_{u,v,w} \equiv \cos \theta_{u,v,w}$ and 
$s_{u,v,w} \equiv \sin \theta_{u,v,w}$, and the two new phases $\phi_{2,3}$ 
($s_{\phi_2,\phi_3} \equiv \sin \phi_{2,3}$) 
are the new parameters which need to be fitted. 
The label $\phi_1$ is reserved for the standard CKM3 phase appearing in (\ref{CKM3}). 
In this parameterization, all matrix elements will now depend on the new parameters, 
for example, the matrix element $V_{ud}$ will have the form
\begin{equation}
V_{ud} = c_w \left(1-\frac{\lambda^2}{2}-\frac{\lambda^4}{8}\right)-e^{i \phi_3}s_w 
(\lambda s_v e^{-i \phi_2}  + A \lambda^3 ( \rho - i \eta ) s_u c_v )\,.
\nonumber 
\end{equation}

In order to estimate CPV phenomena with the 4th generation quarks, 
first we have to determine elements of the
new $4 \times 4$ quark mixing matrix $V_{\rm CKM4}$, (\ref{CKM4def}), 
which essentially means to do a
fitting of the 4th generation parameters
\begin{equation}
m_{b'}, \, m_{t'}, \, s_u, \, s_v, \, s_w, \, s_{\phi_2},\,  s_{\phi_3} \,.
\label{5}
\end{equation}
We shall perform the fit of these parameters 
by analyzing $K^0-\bar{K}^0$, $D^0-\bar{D}^0$ and 
$B_{d,s}^0 - \bar{B}_{d,s}^0$ mixings,
and estimating the decays $K^+ \to \pi^+ \nu\overline{\nu}$ and  $B  \to X_s \gamma$. 
The fit to the 
new measurement of $\sin 2 \beta_{\psi K}$ is added too.  
Moreover, in this analysis we are following 
strict requirement of the unitarity condition
of the new matrix (\ref{CKM4def}) at the expense 
of slight unitarity breaking of the CKM3 matrix. 
This, together with the independently measured CKM3 
matrix elements \cite{Amsler:2008zzb}, 
will give us additional constraints on the parameters of 
the $V_{4 \times 4}$ quark mixing matrix (\ref{4}-\ref{14}).

%%%%%%%%%%%%%%%%%%%%%%%%%%%%%%%%%%%%%%%%%%%%%%%%%%%%%%%%%%%%%

\subsection{Definition of our fitting procedure}

The fit is performed by the CERN fortran code called {\it \bf  Minuit} \cite{MINUIT}. 
It minimizes the multiparameter function 
which is defined as a sum of various $\chi^2$ 's between the fitted expression and the data:
\begin{eqnarray}
\chi^2 (\alpha) = \sum_{i}%^{n_{\rm constraints}}
 \frac{ (th(\alpha)_i - exp_i )^2}{(\Delta th_i)^2 + (\Delta exp_i^2)}\,,
 \label{chis}
\end{eqnarray}
where $th(\alpha)_i$ defines the 4th generation model parameter dependent predictions 
of a given constraint $i$, and $exp$ 
represents the measured values. $\Delta th_i$ is the uncertainty of prediction 
$th_i$ and $\Delta exp_i$ is the uncertainty of the individual measurement $exp_i$.
$\alpha$ is the vector of free parameters being fitted, in our case 
$\alpha =(s_u,s_v,s_w, s_{\phi_2}, s_{\phi_3})$.
The $\chi^2$ will in addition depend on the masses of the $b'$ and $t'$ quarks. 
The fit is performed by varying $m_{b'}$ and $m_{t'}$, 
in such a way that the constraint from the electroweak precision measurements
is fulfilled, assuming $m_H = 115$ GeV(\ref{EWmass1}):
\begin{equation}
m_{b'} = m_{t'} - 55\, {\rm GeV}\,.
\label{EWmass2}
\end{equation}
Much larger mass splitting would require more tuning in the canceling 
contributions to the EW $T$-parameter.\\
A remark is in order: due to the complexity of the expressions, 
the uncertainties of the theoretical predictions 
are not taken into the fit. 

%%%%%%%%%%%%%%%%%%%%%%%%%%%%%%%%%%%%%%%%%%%%%%%%%%%%%%%%%%%%%

\subsection{Fitting different measured processes in the model with four generations}

\vskip 2mm
\noindent
{\bf 2.a.} Check of the unitarity of the CKM3 matrix in SM3

First, we use the unitarity bound on the CKM3 parameters 
coming from the independent measurements. These are:
\begin{eqnarray}
&& |V_{ud}|^2 + |V_{us}|^2 + |V_{ub}|^2 = 0.9999 \pm 0.0011 \quad ({\rm 1st \; row} )\,,
\nonumber\\
&& |V_{cd}|^2 + |V_{cs}|^2 + |V_{cb}|^2 = 1.136 \pm 0.125 \quad ({\rm 2nd \; row} )\,,
\nonumber\\
&& |V_{ud}|^2 + |V_{cd}|^2 + |V_{td}|^2 = 1.002 \pm 0.005 \quad ({\rm 1st \; column} )\,,
\nonumber\\
&& |V_{us}|^2 + |V_{cs}|^2 + |V_{ts}|^2 = 1.134 \pm 0.125 \quad ({\rm 2nd \; column} )\,.
\label{V3S}
\end{eqnarray}
Also, the six CKM3 matrix elements are measured independently, 
\begin{eqnarray}
|V_{ud}| = 0.97418 \pm 0.00027 \,,
\nonumber\\
|V_{us}| = 0.2255 \pm 0.0019 \,,
\nonumber\\
|V_{cd}| = 0.230 \pm 0.011\,, 
\nonumber\\
|V_{cs}| = 1.04 \pm 0.06 \,,
\nonumber\\
|V_{cb}| = (41.2 \pm 1.1 ) \times 10^{-3} \,,
\nonumber\\
|V_{ub}| = (3.93 \pm 0.36 ) \times 10^{-3}\,, 
\label{V3}
\end{eqnarray}
which give us additional six constraints on the unknown parameters. 

\vskip 2mm

\noindent
{\bf 2.b.} $K^0 - \bar{K}^0$ mixing

We modify expressions for the $\Delta M_K$, $\epsilon_K$ and $\epsilon^{\prime}/\epsilon$ 
ratio in the 
model with the fourth generation in order to limit the elements $V_{t'd}$ and $V_{t's}$.  
The experimental values are:
\begin{eqnarray}
\Delta M_K = (3.483 \pm 0.006)\cdot 10^{-15}\,,\\
\epsilon_K \equiv |\epsilon|= (2.229 \pm 0.012)\cdot 10^{-3}\,,\\ 
{\rm Re}(\epsilon^{\prime}/\epsilon) = (1.63 \pm 0.26) \cdot 10^{-3}\,.
\end{eqnarray}

\vskip 2mm

\noindent
{\bf 2.c.} $D^0 - \bar{D}^0$ mixing

An expression for the $x_{D}$ is used in order to determine $V_{c b'}$ and $V_{u b'}$. 
From the experiment we have:
\begin{eqnarray}
x_{D} = 0.776 \pm 0.008\,.
\label{DD} 
\end{eqnarray}

\vskip 2mm

\noindent
{\bf 2.d.} $B^0_{d,s} - \bar{B}^0_{d,s}$ mixings

We need the $x_{B_d}$ and $x_{B_s}$ mixing parameters in order 
to bound $V_{t'd}$\,, $V_{t'b}$\,, and $V_{t's}$\,, $V_{t'b}$\,, respectively.  
The measured values are:
\begin{eqnarray}
x_{B_d} = 0.776 \pm 0.008 \,,\\
x_{B_s} = 26.1 \pm 0.5\,.
\end{eqnarray}

\vskip 2mm

\noindent
{\bf 2.e.} $K^+ \to \pi^+ \nu \bar{\nu}$ process

From the branching ratio for $K^+ \to \pi^+ \nu \bar{\nu}$  
we confine $V_{t'd}$ and $V_{t's}$. Recent experiments give:
\begin{eqnarray}
BR(K^+ \to \pi^+ \nu \bar{\nu}) = (1.5 \pm 1.3 ) \times 10^{-10}\,.
\end{eqnarray}

\vskip 2mm

\noindent
{\bf 2.f.} $B \to  X_s \gamma$ process

To find $V_{t'b}$ we make also use of the branching ratio for $B \to X_s \gamma$ and employ 
\begin{eqnarray}
BR(B \to X_s \gamma) = (3.55 \pm 0.26 ) \times 10^{-4}\,.
\end{eqnarray}

\vskip 2mm

\noindent
{\bf 2.g.} $\sin 2 \beta$ from $B \to J/\psi K$

In the SM3, the best measurement of the $\sin 2 \beta$ comes from 
$B \to J/\psi K$ decay, giving
\begin{equation}
\beta \equiv {\rm arg} \left ( - \frac{V_{cb}^* V_{cd}}{V_{tb}^* V_{td}} \right )\,.
\label{beta}
\end{equation}
In the model with the fourth generation, the $\sin 2 \beta$ can get modified with 
a new phase, i.e. $\sin 2 \beta  \to \sin (2 \beta - 2 \theta)$. 
So, we need to manipulate the expression for $\theta$  
in order to determine $V_{t'b}$ and $V_{t'd}$ from the experimental data,  
\begin{eqnarray}
\sin 2 \beta = 0.681 \pm 0.025 \,.
\label{2beta}
\end{eqnarray}
It is important to note that in all above processes 
the mass $m_{t'}$ appears explicitely in the fit, 
except for $D^0 - \bar{D}^0$ mixing, which depends on the $m_{b'}$ mass (\ref{DD}).

The analytic formulae for the processes are taken from various papers: 
for kaon mixing and decays from 
\cite{Hattori:1999ap}; for $D^0 - \bar D^0$ mixing from
\cite{Golowich:2007ka}; for $B^0_{d,s}-\bar B^0_{d,s}$ mixings from 
\cite{Buras:1990fn}; 
 for $B \to X_s \gamma$ from 
\cite{Greub:1997hf}. 

Various loop-induced processes depend on different Inami-Lim functions \cite{InamiLim}. 
The inclusion of the 4th generation quarks 
in the loops brings additional Inami-Lim functions depending now on $m_{t',b'}/M_W$ 
and the products of the new CKM4 matrix elements $\lambda_{t'}^{bd}$, $\lambda_{t'}^{bs}$, 
$\lambda_{t'}^{sd}$ (and similarly for $b'$), where
\begin{equation}
\lambda_{k}^{lm} = V_{kl}^{\ast} V_{km}\,, 
\label{lambda}
\end{equation}
see for example the analysis in \cite{Yanir:2002cq}. 

All QCD lattice parameters above are taken from the averages in 
\cite{Lubicz:2008am}; see also \cite{Buras:2009pj}. 

We comment here that as was pointed out in \cite{Lunghi:2008aa},
there is tension between the data on $B_d-$ mixing 
and $\sin(2\beta)$ and the theoretical predictions for SM4,
based on the lattice QCD calculations of ref. \cite{Antonio:2007pb}. This 
may signal new physics, such as SM4. Therefore, it is very
important to obtain definite results for the parameters
calculated in lattice QCD. 

In addition, we take into account the findings of two recent studies 
\cite{Bobrowski:2009ng,Chanowitz:2009mz} on 
the 4th generation mixing with the standard 
three quark families. In the first paper \cite{Bobrowski:2009ng}, 
the authors perform similar fit 
like our, by using experimental constraints coming from the measured  
CKM3 matrix elements and FCNC processes 
($K-, D-, B_d-, B_s-$ mixings and the decay $b \to s \gamma$) and asumming 
the unitarity of the new $V_{4 \times 4}$ matrix. 
As it can be seen from above, we have extended the fit adding more FCNC constraints, 
but our results closely follow 
the findings of \cite{Bobrowski:2009ng}, in a sense that the large mixing between 
3rd and 4th generation is allowed for 
some range of the five-dimensional fitting space $\alpha$. 
However, the analysis of second paper \cite{Chanowitz:2009mz} have shown 
that such a large mixing between 
third and the fourth generation, larger than the Cabibbo mixing of the first two families, 
is excluded by the electroweak precision data. 
Therefore, in addition, we apply the EW precision data constraint from 
\cite{Chanowitz:2009mz}, which implies that maximum of 
$\sin \theta_{34} = \sin \theta_u$ must be in the following range  
\begin{eqnarray}
{\rm max}\;({\sin \theta_u}) = 
\left\{
\begin{array}{cl}
0.35 \pm 0.001, & {\rm for} \;\; m_{t'}=300\;\;{\rm GeV}\,,\\ 
0.11 \pm 0.10, & {\rm for} \;\; m_{t'}=1000\;\;{\rm GeV}\,,
\end{array}
\right.
\label{sinT}
\end{eqnarray}
(for other values and for more explanations, see Table 3 in 
\cite{Chanowitz:2009mz}).    
Here, the lower bound for large $m_{t'}$ masses is enlarged, 
%in comparison to those shown in Table 3 of \cite{Chanowitz:2009mz}, 
due to the unreliable perturbation theory applied for the EW fits 
at such large energies (see discussion in \cite{Chanowitz:2009mz}). 

Applying all the constraints discussed above, 
we obtain the results presented in the next subsection. 

%%%%%%%%%%%%%%%%%%%%%%%%%%%%%%%%%%%%%%%%%%%%%%%%%%%%%%%%%%%%%

\subsection{The results of the fitting procedure}

Here are the results for the fitted values of the vector $\alpha = (s_u, s_v, s_w,
s_{\phi_2}, s_{\phi_3})$ depending on the 4th generation quark masses. 
Since we have just one place where the $m_{b'}$ mass enters, ({\bf 2.c.}) from above, 
the quark mass dependence comes mainly from the $m_{t'}$. 

The experimental constraints on the $m_{t'}$ and $m_{b'}$ masses are 
\cite{Amsler:2008zzb,Aaltonen:2007je}:
\begin{eqnarray}
m_{b'} &>& ( 46 - 199 ) \; {\rm GeV}\,,
\\
m_{t'} &>& 256  \; {\rm GeV}\,.
\end{eqnarray} 
Therefore, we scan the $m_{t'}$ in the range of 
\begin{equation}
300 \, {\rm GeV} < m_{t'} < 1000 \, {\rm GeV}\,,
\end{equation}
and take care about the EW precision data limit on 
$m_{b'}$ and $m_{t'}$ mass difference, eq. (\ref{EWmass2}). 

It is important to note that in models with the light Higgs, 
there is an unitary bound  on the masses of the fourth generation quarks which 
amounts to $m_{b',t'} \leq 550$ GeV. If the Higgs boson is heavy ($m_H \geq 500$ GeV), 
the above perturbative limit does not hold, and the masses of 
the fourth family can be larger \cite{Holdom:2009rf,Holdom:2006mr}. 

The quality of the fit is given by the minimal $\chi^2/d.o.f$, where $d.o.f$ is 
the number of the constraints minus 
the number of the fitted parameters. The best fit is when $\chi_{\rm min}^2/d.o.f \approx 1$.
For the numbers given below, $d.o.f = 13$.  

\begin{table}[t]
\begin{center}
\begin{tabular}{|c|| c|| c|}
\hline
$m_{t'}({\rm GeV})$ & $|\sin \theta_u|$ & $\chi_{\rm min}^2/d.o.f$ \\
\hline
%300   &  0.30  & 12.1 \\
300   &  $0.25 \pm 0.04  $  & 0.85 \\
350   &  $0.13 \pm 0.03  $  & 0.98 \\
400   &  $0.10 \pm 0.02  $  & 0.84 \\
450   &  $0.10 \pm 0.04 $  & 0.79 \\
500   &  $0.10 \pm 0.04  $  & 0.80 \\
600   &  $0.11 \pm 0.03  $  & 0.93 \\
700   &  $0.11 \pm 0.02  $  & 1.17 \\
800   &  $0.11 \pm 0.02  $  & 1.45 \\
900   &  $0.11 \pm 0.02  $  & 1.76 \\
1000  &  $0.11 \pm 0.02  $  & 2.07 \\
\hline
\end{tabular}
\end{center}
\caption{Results of our fit on the mixing between 
the third and the fourth generation obtained including the EW constraints from \cite{Chanowitz:2009mz}.}
\end{table}

\begin{table}
\begin{center}
\begin{tabular}{|c|| c c c c c|}
\hline
$m_{t'}({\rm GeV})$ & 300 & 400 & 500 & 600 & 700  \\
\hline
$\sin\theta_u$ & $0.25\pm 0.004$ & $0.10\pm 0.02 $ & $0.10\pm 0.004$ & $0.11\pm 0.03 $ & $0.11 \pm 0.02 $\\ 
$\sin\theta_v$ & $0.010\pm0.003$ & $0.029\pm0.001$ & $0.034\pm0.001$ & $0.033\pm0.008$ & $0.031\pm 0.005$\\  
$\sin\theta_w$ & $0.002\pm0.001$ & $0.016\pm0.002$ & $0.015\pm0.001$ & $0.014\pm0.001$ & $0.012\pm 0.001$\\  
$\sin{\phi_2}$ & $ -0.4\pm 0.4 $ & $0.97\pm 0.01 $ & $0.947\pm0.002$ & $0.91 \pm0.02 $ & $0.89 \pm 0.04 $\\ 
$\sin{\phi_3}$ & $ {\phantom{-}}0.2\pm 0.3  $ & $0.99\pm 0.02 $ & $0.987\pm0.001$ & $0.96 \pm0.03 $ & $0.95 \pm 0.03 $\\  
\hline
\end{tabular}
\end{center}
\caption{Final results for the 4th generation parameters obtained with the acceptable quality fit.}
\end{table}

\begin{table}
\begin{center}
\begin{tabular}{|c|| c c c c c|}
\hline
$m_{t'}({\rm GeV})$ & 300 & 400 & 500 & 600 & 700  \\
\hline 
$|V_{t b }|$  &  $0.964\pm0.010$ & $0.993\pm0.003$ & $0.993\pm0.001$ & $0.992\pm0.003$ & $0.992\pm0.003$\\ 
$|V_{t b'}|$  &  $0.258\pm0.037$ & $0.107\pm0.022$ & $0.106\pm0.004$ & $0.115\pm0.025$ & $0.118\pm0.028$\\
$|V_{t'b }|$  &  $0.259\pm0.037$ & $0.108\pm0.022$ & $0.106\pm0.004$ & $0.115\pm0.025$ & $0.118\pm0.028$\\
$|V_{t'b'}|$  &  $0.965\pm0.010$ & $0.993\pm0.003$ & $0.994\pm0.001$ & $0.992\pm0.003$ & $0.992\pm0.003$\\
$|V_{t'd }|$  &  $0.002\pm0.001$ & $0.009\pm0.002$ & $0.008\pm0.001$ & $0.006\pm0.001$ & $0.006\pm0.001$\\
$|V_{t's }|$  &  $0.005\pm0.005$ & $0.031\pm0.001$ & $0.034\pm0.007$ & $0.034\pm0.009$ & $0.031\pm0.005$\\
$|V_{u b'}|$  &  $0.002\pm0.001$ & $0.016\pm0.002$ & $0.016\pm0.004$ & $0.014\pm0.001$ & $0.013\pm0.002$\\
$|V_{c b'}|$  &  $0.010\pm0.003$ & $0.030\pm0.001$ & $0.033\pm0.002$ & $0.034\pm0.009$ & $0.031\pm0.005$\\
\hline
\end{tabular}
\end{center}
\caption{Predictions for the selected $V_{\rm CKM}$ matrix elements using the best fit from Table 2.}
\end{table}

The following results of the fitting procedure are of a special importance:
\\
{\bf i)} $m_{t'} \sim [300 - 600] \, {\rm GeV}$ region is preferred by the $\chi^2$ scan, 
i.e. $\chi^2_{\rm min}/d.o.f  \approx 1$.
The larger $m_{t'}$ masses do not produce a good fit, as one can see from the Table 1.
\\
{\bf ii)} the best fits with $\chi^2_{\rm min}/d.o.f  \approx 1$ for $m_{t'} > 600$ GeV give too large 
$s_u = \sin \theta_u$ mixing angle which is 
excluded by the EW precision data. On the other hand, 
the allowed values for $s_u$ are obtained
with the bad fit, with $\chi^2_{\rm min}/d.o.f. > 1$, see Table 1. 
\\
{\bf iii)} in addition, we test the predictions for all quantities entering 
the fit using the new fitted parameters,
in a way that we look for 
the 'pull'(= (data central value - predicted value)/(data error) ) of the data.
So, although the fit for $m_{t'}=700$ GeV has $\chi^2_{\rm min}/d.o.f$ 
slightly larger than 1, we have decided to keep this fit,
since the predictions with this mass of $m_{t'}$ nicely match with the data.

Having in mind all the facts above, we conclude that our best fits are obtained for
$300 \leq m_{t'} \leq 700$, with the fitted parameters given in Table 2., while  
the selected $V_{\rm CKM}$ matrix elements are presented in Table 3.

The final results at 95\% confidence level of the complete 
$4\times 4$ fitted matrices are given below:
\begin{eqnarray}
V_{\rm CKM4}(m_{t'}=300\, {\rm GeV}) &=&
\left(
\begin{array}{llll}
 0.9742 & 0.2257 & 0.0035 e^{-68.9^{\circ} i} & 0.0018 e^{-12.4^{\circ} i} \\
 -0.2255  & 0.9732 & 0.0414  & 0.0102 e^{29.8^{\circ} i} \\
0.0086 e^{-24.1^{\circ} i} & -0.0416 e^{0.7^{\circ} i} & 0.9649 & 0.2589 \\
-0.0019 e^{18.9^{\circ} i} & 0.0052 e^{69.3^{\circ} i} & -0.2591  & 0.9658
\end{array}
\right)\,,
\nonumber \\
\label{300}
\end{eqnarray}
\begin{eqnarray}
V_{\rm CKM4}(m_{t'}=400\, {\rm GeV}) &=&
\left(
\begin{array}{llll}
 0.9740 & 0.2256 & 0.0036 e^{-68.9^{\circ} i} & 0.0164 e^{-87.4^{\circ} i} \\
 -0.2259 & 0.9728  & 0.0414  & 0.0290 e^{-76.1^{\circ} i} \\
 0.0092 e^{-27.7^{\circ} i} & - 0.0414  & 0.9932   & 0.1079 \\
 - 0.0091 e^{89.9^{\circ} i} & 0.0310 e^{-94.6^{\circ} i} & -0.1082  & 0.9935
\end{array}
\right)\,,
\nonumber \\
\label{400}
\end{eqnarray}
\begin{eqnarray}
V_{\rm CKM4}(m_{t'}=500\, {\rm GeV}) &=&
\left(
\begin{array}{llll}
 0.9740 & 0.2256 & 0.0035 e^{-68.9^{\circ} i} & 0.0160 e^{-81.1^{\circ} i} \\
 -0.2259  & 0.9726  & 0.0414 & 0.0329 e^{-71.8^{\circ} i} \\
0.0083 e^{-27.1^{\circ} i} & -0.0416 e^{6.0^{\circ} i} & 0.9934  & 0.1059 \\
-0.0080 e^{83.2^{\circ} i} & 0.0344 e^{-100.4^{\circ} i} & -0.1062 e^{0.7^{\circ} i} & 0.9937
\end{array}
\right)\,,
\nonumber \\
\label{500}
\end{eqnarray}
\begin{eqnarray}
V_{\rm CKM4}(m_{t'}=600\, {\rm GeV}) &=&
\left(
\begin{array}{llll}
 0.9741 & 0.2256 & 0.0035 e^{-68.9^{\circ} i} & 0.0140 e^{-75.4^{\circ} i} \\
- 0.2258  & 0.9726 & 0.0414  & 0.0339 e^{-66.0^{\circ} i} \\
0.0089 e^{-26.2^{\circ} i} & -0.0423 e^{6.3^{\circ} i} & 0.9924  & 0.1149 \\
-0.0058 e^{77.9^{\circ} i} & 0.0343 e^{-105.9^{\circ} i} & -0.1155 e^{0.7^{\circ} i} & 0.9926
\end{array}
\right)\,,
\nonumber \\
\label{600}
\end{eqnarray}
\begin{eqnarray}
V_{\rm CKM4}(m_{t'}=700\, {\rm GeV}) &=&
\left(
\begin{array}{llll}
 0.9741 & 0.2256 & 0.0035 e^{-68.9^{\circ} i} & 0.0130 e^{-72.9^{\circ} i} \\
-0.2258  & 0.9727  & 0.0414 & 0.0309 e^{-62.9^{\circ} i} \\
0.0088 e^{-26.2^{\circ} i} & -0.0423 e^{5.8^{\circ} i} & 0.9920  & 0.1179 \\
-0.0056 e^{74.7^{\circ} i} & 0.0309 e^{-108.1^{\circ} i} & -0.1185 e^{0.6^{\circ} i} & 0.9924
\end{array}
\right)\,.
\nonumber \\
\label{700}
\end{eqnarray}
\vskip 2mm
%\hline
%\vskip 2mm
%\noindent
Note, that the fitted parameters show small 4th generation mass dependence 
in the preferable range of $m_{t'}$, excluding the fitted $4 \times 4$ matrix
at $m_{t'} = 300$ GeV, (\ref{300}).

From the above matrices we can see that the fit exhibits constraint
$|V_{tb}| > 0.96$, which is much stronger than 
the limit $|V_{tb}| > 0.74$ following from the single top quark production cross section measurement 
\cite{Amsler:2008zzb}. 

Comparing our results with those existing in the literature 
\cite{Bobrowski:2009ng,Arhrib:2006pm,Arhrib:2009ew}, 
we can deduce that our fit, under the conditions 
specified in Sec. 2.2, excludes large mixing between 4th and the first three generations. 
Our matrix elements $|V_{ub'}|, |V_{cb'}|, |V_{tb'}|$ from Table 3. are significantly
smaller (up to six times for $|V_{tb'}|$) with respect to the same elements obtained by 
the conservative bound in \cite{Bobrowski:2009ng} and in \cite{Herrera:2008yf}. 
This is a direct consequence of the applied 
EW constraint on $\sin \theta_u$, (\ref{sinT}), since otherwise, 
as already mentioned at the end of Sec. 2.2, the 
somewhat larger mixing between the third and the fourth generation, relative to 
the bound from eq.(\ref{sinT}), is obtained. 
In \cite{Yanir:2002cq}, the  mixing is bounded to $\sin \theta_{34} \le 0.14$.

Considering phases of CKM4, in our approach they
are strongly depending on the $m_{t'}$ mass, oscillating widely, 
as they do in \cite{Bobrowski:2009ng}.
In \cite{Yanir:2002cq}, as well as in \cite{Bobrowski:2009ng,Herrera:2008yf},
the fits are performed
under the assumption that the phases are free and run between $0$ and $2 \pi$.
However, in our global and unique fit, which generates matrices (\ref{300} - \ref{700}), 
the phases are also subject of the fitting procedure. Therefore, 
the complex interplay between all fitting parameters 
can significantly influence the final allowed parameter values of the matrix elements. 

Although, the standard CKM3 matrix elements, 
as a part of $V_{\rm CKM4}$ were fitted, in our fit 
their values (\ref{300} - \ref{700}) do not contradict 
the global CKM3 fit from \cite{Amsler:2008zzb}.
This is especially true for the less constrained elements 
like $V_{td}$ and $V_{ts}$.

The obtained fourth generation parameter values (\ref{300} - \ref{700}) 
will be used in the calculation of the 
rare decay branching ratios and CP partial rate asymmetry in the next sections. 

%%%%%%%%%%%%%%%%%%%%%%%%%%%%%%%%%%%%%%%%%%%%%%%%%%%%%%%%%%%%%

\section{Rare processes involving the fourth generation}
\setcounter{equation}{0}   

\begin{figure}
\begin{center}
\subfigure[$\;b'$ decays]{\includegraphics[scale=0.6]{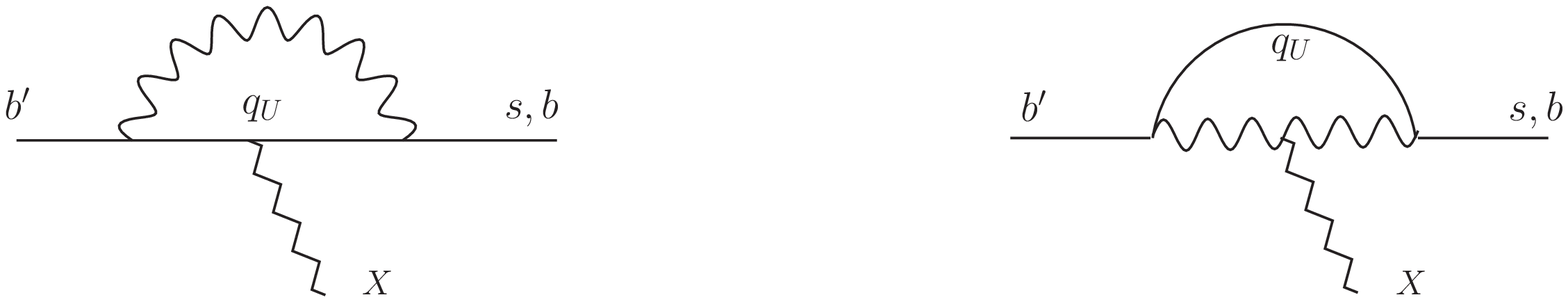}}
\hspace{0.7cm}
\subfigure[$\;t'$ decays]{\includegraphics[scale=0.6]{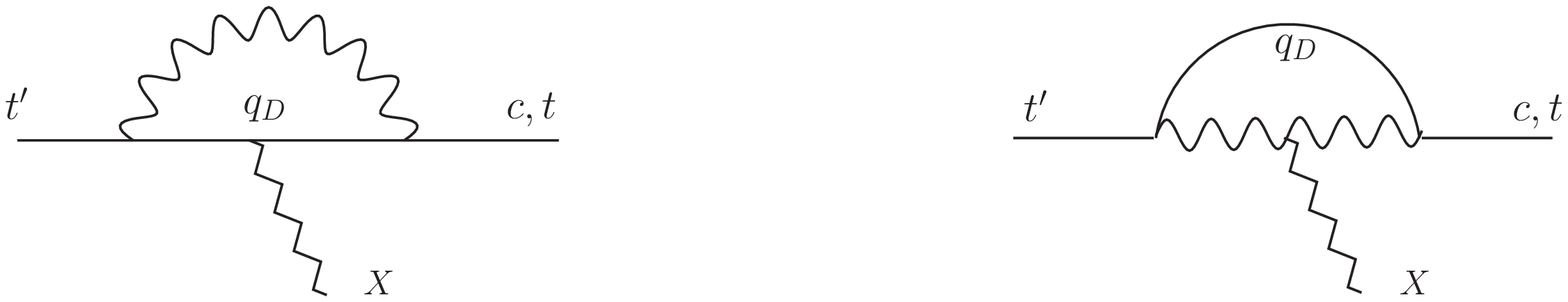}}
\caption{\em Generic diagrams for FCNC decays of the 4th generation quarks. 
$X$ denotes possible decays to $X =  H, Z, \gamma, g$ and quarks running in 
the loops are $q_U = \{ u,c,t,t'\}$ and $q_D = \{ d,s,b,b' \}$.}
\label{fig-decays}
\end{center}
\end{figure}

We analyze FCNC decay processes of the fourth generation quarks, 
in particular of $t' \to (c,t)X$, and $b' \to (s,b)X$ with $(X=H, Z, \gamma, g)$, 
arising from the generic one-loop diagrams given in Figs.\ref{fig-decays}. 
We also study the influence of the 4th generation FCNC model to the ordinary 
top quark rare decays: $t \to c X$.

The rare FCNC processes of the above type have been extensively studied in 
the context of various extensions of SM. 
We base our study on the explicit analytical expressions on 
$Q \to q (Z,\gamma, g)$ given in 
\cite{Eilam:1990zc}; with $Q=(t,t',b')$ and $q=(c,(t,c),(b,s))$, respectively. 
Checks for $Q \to q (\gamma, g)$ 
decays are performed using expressions from \cite{AguilarSaavedra:2002ns}. 
The $Q \to q H$ decays were considered 
in \cite{Haeri:1988jt}. 

\begin{figure}[t]
\begin{center}
\includegraphics[scale=0.80]{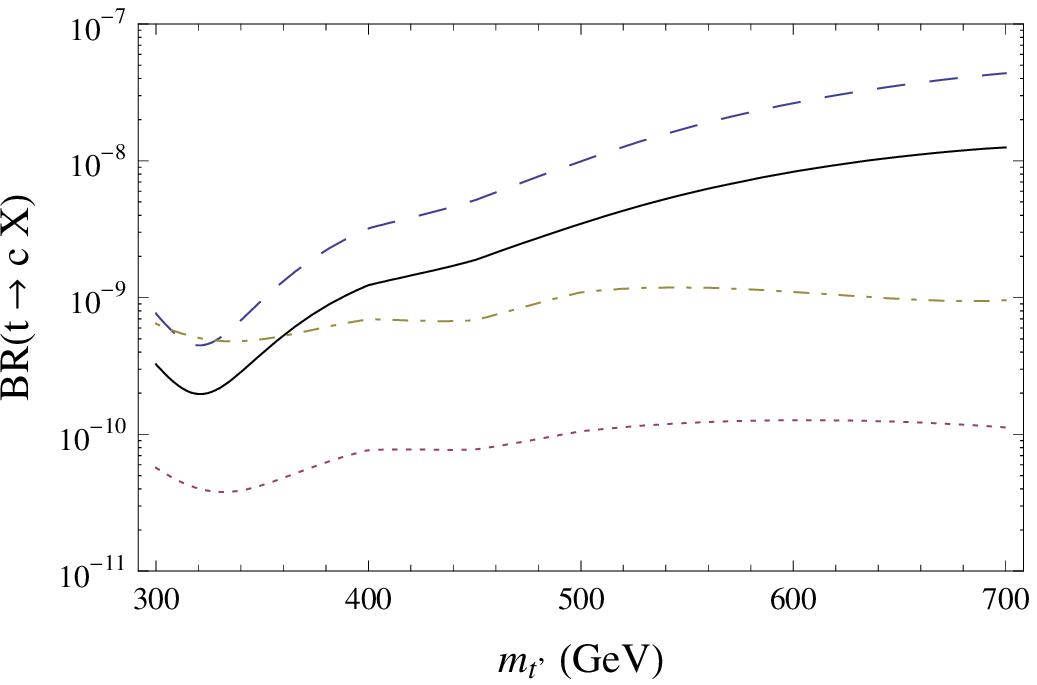}
\caption{\em Branching ratios for rare top decays in the model with 4th generation 
as a function of $m_{t'}$, for  $b'$ of the mass $m_{b'} = m_{t'} - 55$ GeV running in the loops.
$X$ denotes possible decays to 
$X$ = $H$(dashed line), $Z$(solid line), $\gamma$(dotted line), $g$(dashed-dotted line).}
\label{figBRtcX}
\end{center}
\end{figure}

\begin{figure}[h]
\begin{center}
\subfigure[$\;BR(t' \to c X)$]{\includegraphics[scale=0.73]{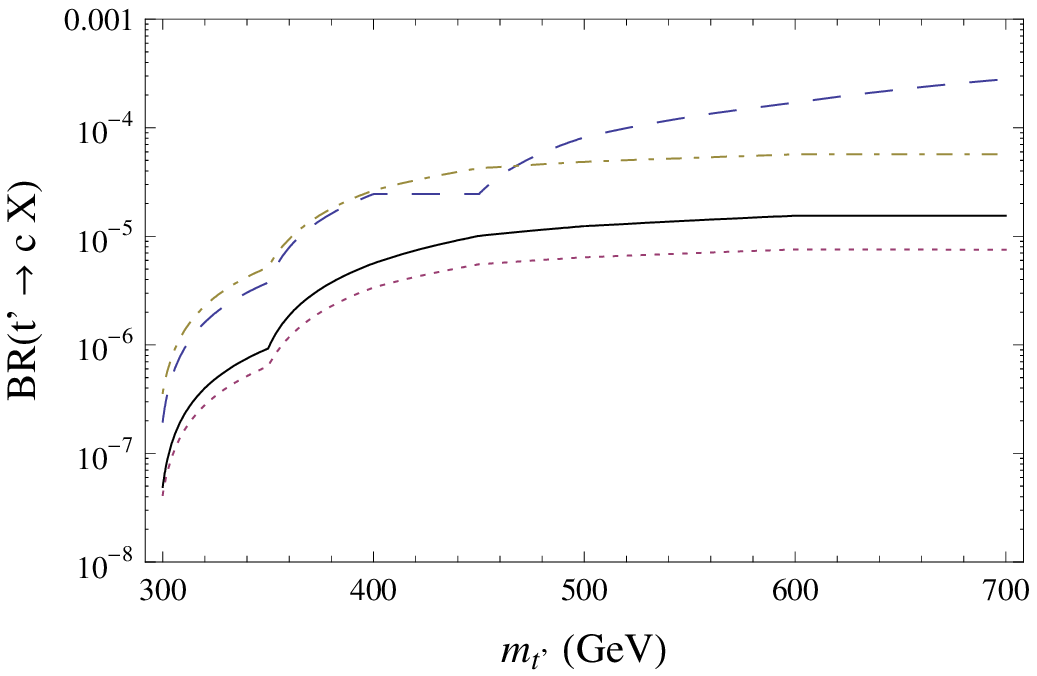}}
\hspace{0.7cm}
\subfigure[$\;BR(t' \to t X)$]{\includegraphics[scale=0.73]{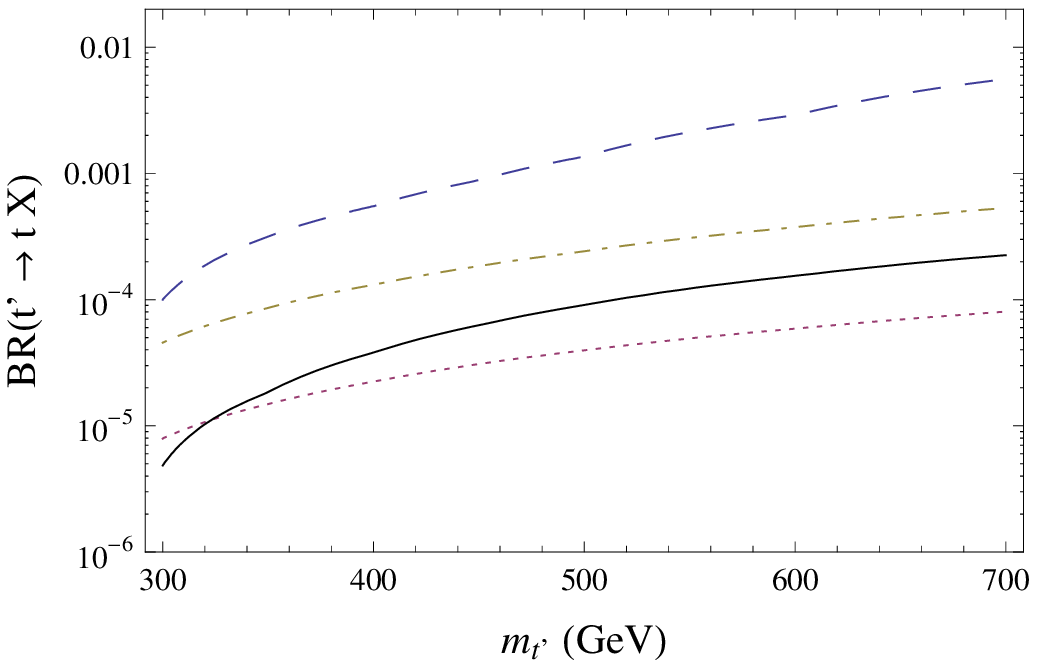}}
\caption{\em Branching ratios of $t' \to (c,t)X$ as a function of $m_{t'} = m_{b'} + 55$ GeV.
$X$ denotes possible decays to 
$X$ = $H$(dashed line), $Z$(solid line), $\gamma$(dotted line), $g$(dashed-dotted line).}
\label{figBRtpX}
\end{center}
\end{figure}

To obtain the branching ratios, the decay amplitudes will be normalized to the widths of 
the decaying quarks. For $t$-quark decays we have 
\begin{eqnarray}
{\rm BR}(t \to c X) = \frac{\Gamma(t \to c X)}{\Gamma(t \to b W)}\,,
\label{BRtc}
\end{eqnarray}
while for $t'$ and $b'$ decays, we will also take into account 
the CKM4-suppressed tree level decays. Therefore, 
\begin{eqnarray}
{\rm BR}(t' \to (c,t)X) &=& \frac{\Gamma(t'\to (c,t)X)}{\Gamma(t' \to bW) + \Gamma(t' \to sW)}\,,
\label{BRtpc}\\
{\rm BR}(b' \to (s,b)X) &=& \frac{\Gamma(b'\to (s,b)X)}{\Gamma(b'\to tW^{(*)})+\Gamma(b'\to cW)}\,,
\label{BRbps}
\end{eqnarray}
where $b' \to t W^{*}$ is effective for $m_{b'} \le 255$ GeV. 
The tree level decays are given by
\begin{eqnarray}
\Gamma(Q\to q W) &=& \frac{G_F M_W^3 x_Q^3}{8\pi\sqrt{2}}
|V_{Qq}|^2\sqrt{\lambda\left(1,(1/x_Q)^2,(x_q/x_Q)^2 \right )} 
\nonumber \\
&& \times \left ( \left (1 - x_q^2/x_Q^2\right )^2 + 1/x_Q^2 \left ( 1 + x_q^2/x_Q^2 \right ) 
- 2/x_Q^4 \right )\,,
\label{Width}
\end{eqnarray}
where $\lambda(x,y,z) = x^2 + y^2 + z^2 - 2 x y - 2 x y - 2 y z$ and $x_i^2 = m_i^2/M_{W}^2$.
\begin{figure}
\begin{center}
\subfigure[$\;BR(b' \to s X)$]{\includegraphics[scale=0.73]{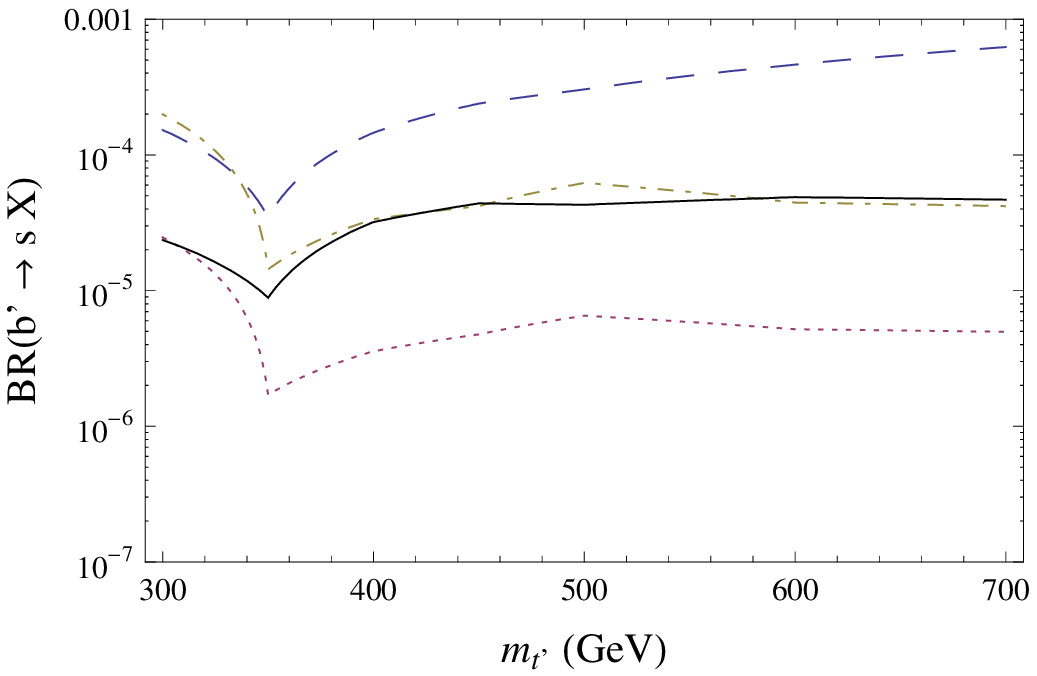}}
\hspace{0.7cm}
\subfigure[$\;BR(b' \to b X)$]{\includegraphics[scale=0.73]{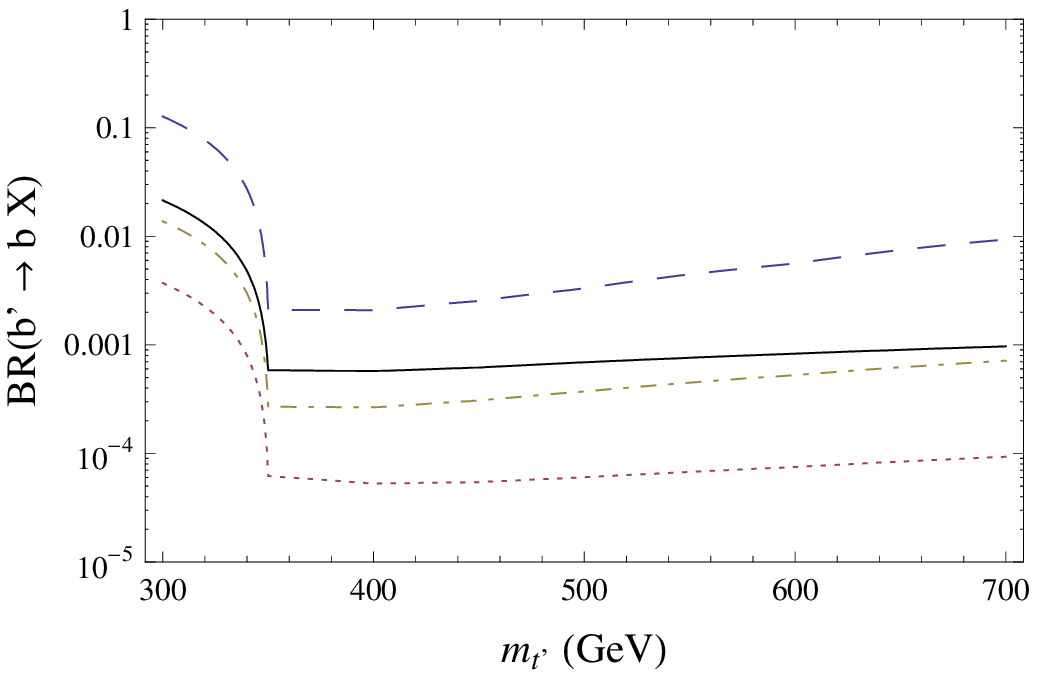}}
\caption{\em Branching ratios of $b' \to (s,b)X$ as a function of $m_{t'} = m_{b'} +  55$ GeV. 
$X$ denotes possible decays to 
$X$ = $H$(dashed line), $Z$(solid line), $\gamma$(dotted line), $g$(dashed-dotted line).} 
\label{figBRbpX}
\end{center}
\end{figure}
The masses running in the loops in Figs.\ref{fig-decays} are taken to be current quark masses, 
while the external masses are considered 
as pole masses. Practically, this makes no 
numerical difference in the calculation, apart for $t \to c X $ decays 
with $b$-quark running in the loops. 
There we take $\overline{m}_b(m_t) = 2.74$ GeV, \cite{AguilarSaavedra:2002ns}, 
and, with our set of parameters and $m_t = 171.2$ GeV, we obtain 
the following SM results: 
\begin{eqnarray}
{\rm BR}_{SM}(t \to c \gamma) &=& 4.4 \cdot 10^{-14}\,,
\qquad\qquad 
{\rm BR}_{SM}(t \to c g) = 3.8 \cdot 10^{-12}\,,
\nonumber\\
{\rm BR}_{SM}(t \to c Z) &=& 1.3 \cdot 10^{-14}\,,
\qquad\qquad 
{\rm BR}_{SM}(t \to c H) = 7.8 \cdot 10^{-15}\,,
\label{tSM}
\end{eqnarray}
comparable with the estimates given in \cite{AguilarSaavedra:2002ns,AguilarSaavedra:2004wm}. 
With the pole masses running in the loops \cite{Eilam:1990zc}, 
the results become an order of magnitude larger.  
The values from eqs. (\ref{tSM}) have to be compared with the largely enhanced BRs  
of $t \to c X$ for 4th generation quarks 
included in the loops, Fig.\ref{figBRtcX}. The foreseen sensitivities for $t \to c X$ channels 
at Tevatron and/or LHC could be sufficient to see these enhanced rates.  

Throughout of the calculation the mass of the Higgs boson is taken to be $m_H = 115$ GeV. 
Our fit favors fourth generation masses slightly larger than the so-called 
unitary bound of $\sim 600$ GeV. In that case the 
concept of light Higgs boson and the elementary scalar Higgs field is no more appropriate, 
since the Goldston boson of the electroweak symmetry breaking would couple very strongly to 
the heavy 4th generation quarks
\cite{Holdom:2009rf,Holdom:2006mr}. Therefore, the results for 
$m_{t',b'} \geq 600$ GeV have to be taken with precaution. 

In the analysis we have also examined the influence of the $W$-boson width to the results. 
The inclusion of the finite width for the $W$-boson propagating through the loops,  
\cite{Eilam:1991yv}, is effective only for the 
$t \to c X$ decays, enhancing BRs by some $10\%$. 
 
Our prediction for $BR(t \to cH)$ given in Fig.\ref{figBRtcX}, contrary to \cite{Arhrib:2006pm},
always dominates over $Z, \gamma, g$ modes, in the whole range of $t'$ mass.
For $t' \to (c,t)X$  (Fig.\ref{figBRtpX}) and $b' \to (s,b)X$  (Fig.\ref{figBRbpX}) 
decay modes general behavior 
is more or less the same,  except that for our global fit, both, 
the gluon and the Higgs modes dominate over Z and photon modes,
apart from the case given in Fig.\ref{figBRbpX}b, where $H$ and $Z$ dominate over 
$g$ and $\gamma$ modes, respectively. In \cite{Arhrib:2006pm}
dominating modes are Z and the decay into gluon, 
which is due to a large difference between 
our CKM4 parameters and the parameters used in \cite{Arhrib:2006pm}. 

%%%%%%%%%%%%%%%%%%%%%%%%%%%%%%%%%%%%%%%%%%%%%%%%%%%%%%%%%%%%

\section{CP violation}
\setcounter{equation}{0}   

The CP partial rate asymmetry, for decays discussed above, is defined as 
\begin{eqnarray}
a_{CP}=\frac{\Gamma(Q \to q X)-\Gamma(\bar Q \to \bar q \bar{X})}
{\Gamma(Q \to q X)+\Gamma(\bar Q\to \bar q \bar X)}\,.
\label{aCP4}
\end{eqnarray}
Since the rates involve at least two amplitudes 
with different CP-conserving strong phases coming from the absorptive parts of the loops, 
while the CP-violating 
weak phases are provided by the phases in $V_{\rm CKM4}$, we expect 
to find CP violation in FCNC decays of 4th generation quarks \cite{Atwood:2000tu}. 
The inclusion of the finite $W$-boson width can enhance 
a CP-asymmetry by enhancing the CP-conserving phases, 
but, since this happens almost equally for $\Gamma(Q \to q X)$ 
and $\Gamma(\bar Q \to \bar q \bar{X})$, the effect appear to be at most of 10\% level. 
\begin{figure}
\begin{center}
\includegraphics[scale=0.80]{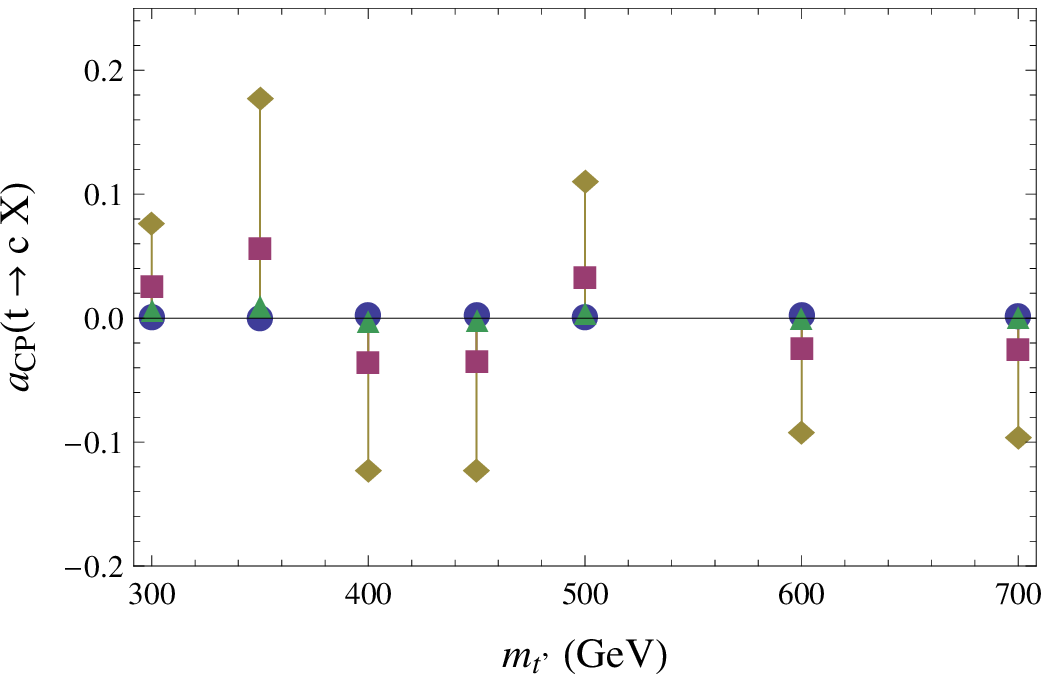}
\caption{\em Fourth generation effect on the $a_{CP}$ of the rare top decays as a function 
of $m_{t'}$, for $b'$ of the mass $m_{b'} = m_{t'} -  55$ GeV running in the loops.
$X$ denotes possible decays to 
$X$ = $H$({\tiny \ding{108}}), $Z$({\tiny \ding{115}}), $\gamma$({\tiny \ding{110}}), 
$g$({\tiny \ding{117}}).} 
\label{figACPtcX}
\end{center}
\end{figure}
\begin{figure}[t]
\begin{center}
\subfigure[$\;a_{CP}(t' \to c X)$]{\includegraphics[scale=0.73]{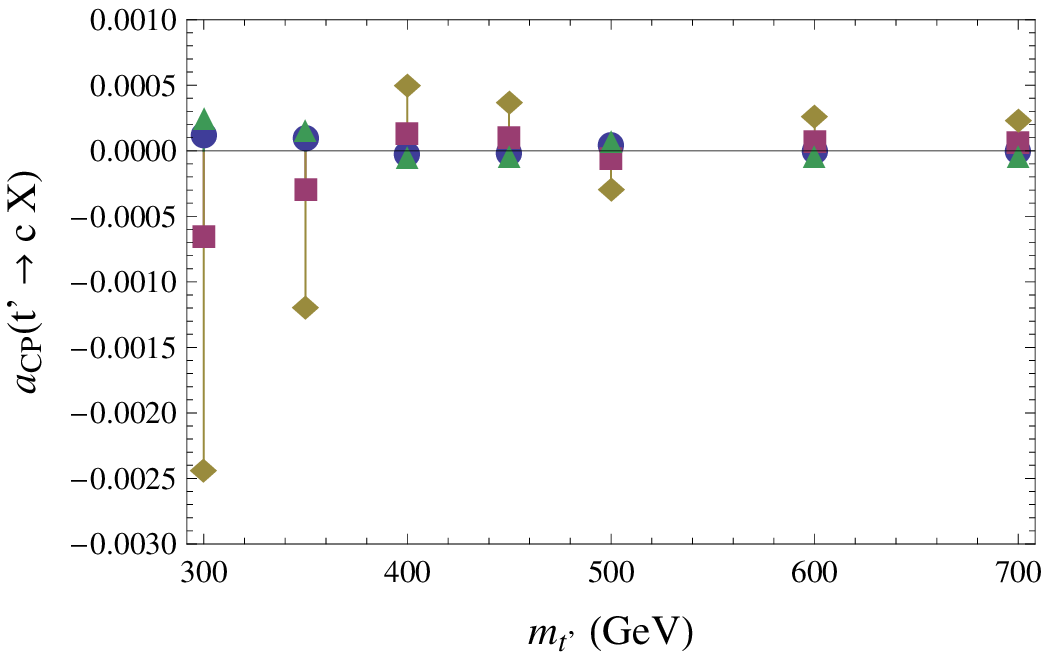}}
\hspace{0.7cm}
\subfigure[$\;a_{CP}(t' \to t X)$]{\includegraphics[scale=0.74]{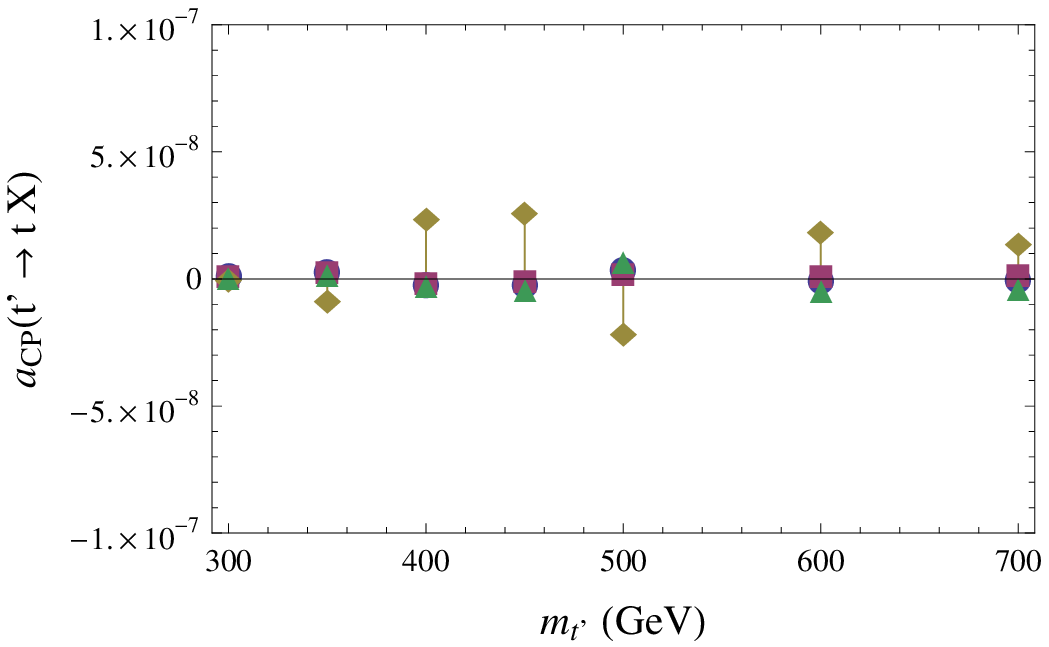}}
\caption{\em CP asymmetries in  $t' \to(c,t)X$ decays as a function of $m_{t'} = m_{b'} +  55$ GeV. 
$X$ denotes possible decays to 
$X$ = $H$({\tiny \ding{108}}), $Z$({\tiny \ding{115}}), $\gamma$({\tiny \ding{110}}), 
$g$({\tiny \ding{117}}).} 
\label{figACPtpX}
\end{center}
\end{figure}
\begin{figure}[t]
\begin{center}
\subfigure[$\; a_{CP}(b' \to s X)$ ]{\includegraphics[scale=0.73]{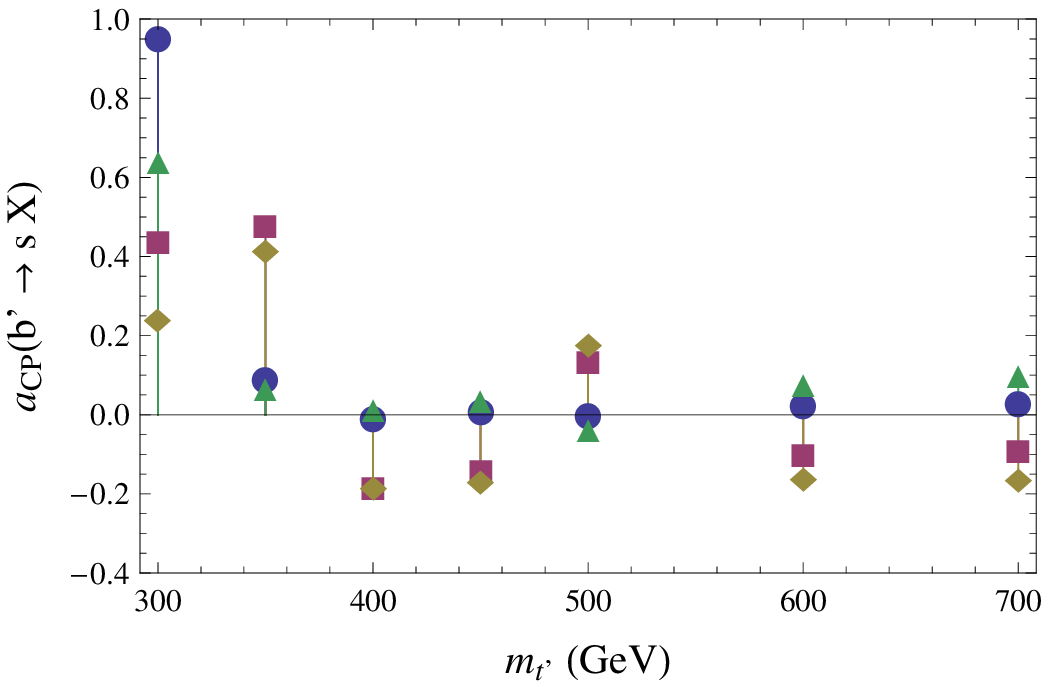}}
\hspace{0.7cm}
\subfigure[$\; a_{CP}(b' \to b X)$ ]{\includegraphics[scale=0.74]{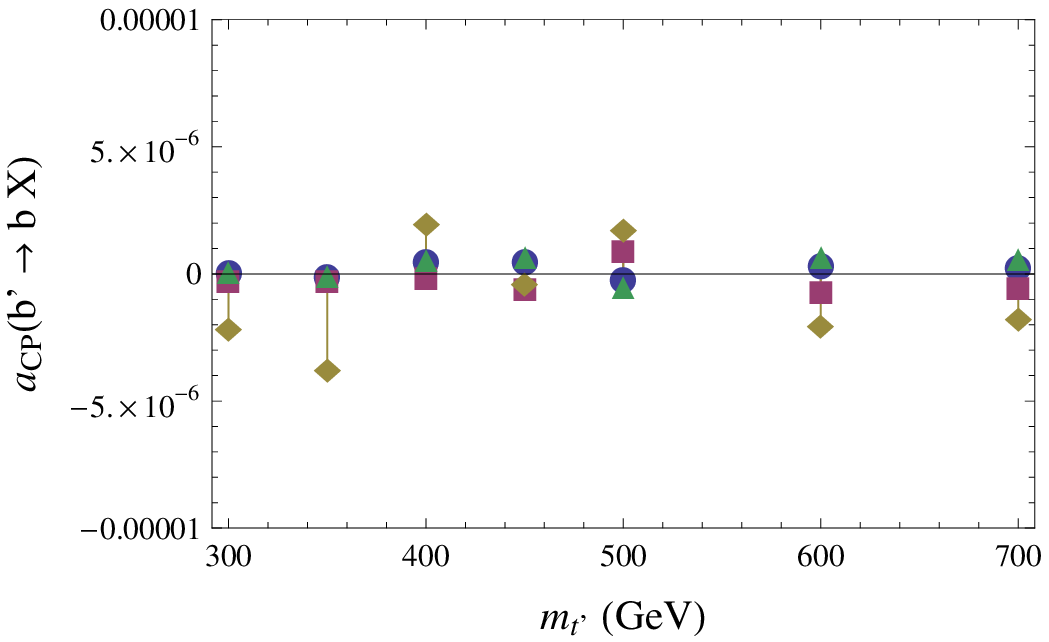}}
\caption{\em CP asymmetries in  $b' \to (s,b)X$ decays as a function of $m_{t'} = m_{b'} +  55$ GeV. 
$X$ denotes possible decays to 
$X$ = $H$({\tiny \ding{108}}), $Z$({\tiny \ding{115}}), $\gamma$({\tiny \ding{110}}), 
$g$({\tiny \ding{117}}).} 
\label{figACPbpX}
\end{center}
\end{figure}
Estimated CP asymmetries, shown in Figs.\ref{figACPtcX}-\ref{figACPbpX} 
for FCNC rare decay modes, 
in general oscillate as a function of $t'$ mass.
In particular, important modes for CPV effects are $b' \to s X$ decays, 
as also noted recently in \cite{Arhrib:2009ew}. 
For $b' \to s(H,Z)$ modes we find very interesting maximal
CP partial rate asymmetry at $m_{t'}=300$ GeV, i.e. $94$\% and $62$\% respectively.
These large numbers occur due to the $tW$ loop threshold at $m_{b'} \simeq 250$ GeV. 
Other two modes, $b' \to s(\gamma,g)$,
produce maximal CPV at $m_{t'}=350$ GeV in 
the amount of $47$\% and $41$\% for $\gamma$ and $g$, respectively, and
they could be important too, Fig.\ref{figACPbpX}a. Maximal CP partial rate 
asymmetry for $t \to cX$ modes occur also at $m_{t'}=350$ GeV for $t \to cg$, 
and it amounts to  $18$\%, Fig.\ref{figACPtcX}. 
For $t' \to (c,t)X$ and $b' \to bX$ modes $a_{CP}$ is very small, 
always bellow 0.25\%, Figs.\ref{figACPtpX},\ref{figACPbpX}b.

At the end, let us discuss some general features of the CP violation within the model with the 4th family. 

Following the analysis of Ref.\cite{delAguila:1997vn}, we calculate 
the strengths $|B_i|$ of CP violation for a fourth family in the chiral 
limit $m_{u,d,c} = 0$. Definitions of the relevant imaginary products in the chiral limit are 
\cite{delAguila:1997vn}:
\begin{eqnarray}
B_1 &\equiv& {\rm Im} V_{cb} V_{t'b}^{\ast} V_{t'b'} V_{c b'}^{\ast} \,,
\label{BiIm1}\\
B_2 &\equiv& {\rm Im} V_{tb} V_{t'b}^{\ast} V_{t'b'} V_{t b'}^{\ast} \,,
\label{BiIm2}\\
B_3 &\equiv& {\rm Im} V_{cb} V_{t b}^{\ast} V_{t b'} V_{c b'}^{\ast} \,.
\label{BiIm3}
\end{eqnarray}
In the paper \cite{delAguila:1997vn} a  rigorous upper bound on $|B_{i}| \le 10^{-2}$ in the model 
with the 4th family was obtained. 
Calculating these quantities explicity for the values of our CKM4 matrix elements 
(\ref{300}-\ref{700}), we obtain the strengths of the CP violation of the order 
\begin{eqnarray}
|B_{1,2,3}| \simeq
\left\{
\begin{array}{cl}
5 \cdot 10^{-5} & {\rm for} \;\; m_{t'}= 300\;\;{\rm GeV},\\ 
10^{-4} & {\rm for} \;\; m_{t'}= [400-700]\;\;{\rm GeV}.\\ 
\end{array}
\right.
\label{B123}
\end{eqnarray}
The area of the unitary quadrangle $A_{bb'}$\,, with the sides 
$V_{ub}V_{ub'}^{\ast}$\,, $V_{cb} V_{cb'}^{\ast}$\,, $V_{tb} V_{tb'}^{\ast}$\,, 
$V_{t'b} V_{t'b'}^{\ast}$\,, describing CPV in the chiral limit, is 
\begin{eqnarray}
A_{bb'} = \frac{1}{4} \{ |B_1 + B_2| + |B_1 + B_3| + |B_2| + |B_3| \}\,,
\label{AB123}
\end{eqnarray}
and with our fitted parameters amounts to 
\begin{eqnarray}
2 A_{bb'} \simeq
\left\{
\begin{array}{cl}
10^{-5} & {\rm for} \;\; m_{t'}= 300\;\;{\rm GeV},\\ 
4 \cdot 10^{-4} & {\rm for} \;\; m_{t'}= [400-700]\;\;{\rm GeV}.\\ 
\end{array}
\right.
\label{Abbp}
\end{eqnarray}
The same values are obtained for the area of the unitary quadrangle 
from eq.(\ref{Jarlskog}), defined by the unitarity relation 
$V_{us}V_{ub}^{\ast}+V_{cs}V_{cb}^{\ast}+V_{ts}V_{tb}^{\ast}+V_{t's}V_{t'b}^{\ast}=0$. 
This has to be compared with the amount of CPV in the three-generation SM given by 
$|{\rm Im} V_{ij} V_{kj}^{\ast} V_{kj} V_{il}^{\ast} |  \le 5 \times 10^{-5}$. 

We see that the measure of the CPV in the 4th generation model is only 
slightly larger than the amount of CPV 
in SM3 and this happens only for larger extra quark masses. 
It seems that extra quarks can give us 
new sources of large CPV phenomena, but, in general, cannot bring significant cumulative 
effect in the strength of CPV (\ref{Abbp}).
Therefore, a huge enhancement in the Jarlskog invariant $J_{234}^{bs}$  (\ref{Jarlskog}), 
in the model with the 
4th family, comes predominantly from the 
large $m_{b'}$ and $m_{t'}$. 

%%%%%%%%%%%%%%%%%%%%%%%%%%%%%%%%%%%%%%%%%%%%%%%%%%%%%%%%%%%%%
%%%%%%%%%%%%%%%%%%%%%%%%%%%%%%%%%%%%%%%%%%%%%%%%%%%%%%%%%%%%

\section{Discussion and Conclusions}
\setcounter{equation}{0}   

In this paper we investigate the CP-violating decay processes involving 
the fourth quark generation and find large CP partial rate asymmetries 
for some of decay modes. 
We achieve that by constructing and employing global unique fit of the unitary CKM4 mass matrix.
Our fit for certain values of the 4th generation quark mixing matrix elements   
for $300 \leq m_{t'} \leq 700$ GeV, produces highly enhanced $a_{CP}$ for $b' \to s $ decay modes.
A dominance of 
$a_{CP}(b^{\prime} \to s\;(H,Z;\,\gamma,g))=\,  (95,62;\,47,41)\%$ at 
$m_{t'}\simeq 300;\,350$ GeV with respect to all other modes is particularly interesting. 

It is important to note here that all quantities appearing in the
4th generation mixing matrix were subject to our fitting procedure, contrary to 
\cite{Yanir:2002cq,Bobrowski:2009ng,Herrera:2008yf}.
So, the phases of $V_{\rm CKM4}$ are fitted too, and, the complex interplay between all fitted parameters 
significantly influences the final fit of the matrix elements (\ref{300}-\ref{700}), 
and therefore the estimated CP partial rate asymmetries as well.

We have inspected FCNC decay processes of the 4th generation quarks, 
$b^{\prime} \to s\,X,\;b^{\prime} \to b\,X,\;t^{\prime} \to c\,X,\;t^{\prime} \to t\,X$,
with $X=H,\,Z,\,\gamma,\, g$, and the top decays $t \to c X$ 
for 4th generation quarks running in the loops.
The branching ratios of these rare top decays gets highly enhanced 
due to the presence of the 4th family quarks. 
Considering first the CPV effects for $t \to c\,X$ modes, 
we have found $|a_{CP}(t \to c g)| \simeq 8-18$ \% at $m_{t'}= 300-700$ GeV; for $t \to c\gamma$ 
mode $a_{CP}$ is always bellow 6\%, 
while for $t \to c(H,Z)$ asymmetries are negligible, Fig.\ref{figACPtcX}.
The $a_{CP}$, as a function of $t'$ mass between 300 and 700 GeV, oscillate for all decay modes, 
Figs.\ref{figACPtcX}-\ref{figACPbpX}.
As already noted, the $b^{\prime} \to s\;(H,Z)$ modes with $95(62)\%$ CP asymmetries at 
$m_{t'} = 300$ GeV dominate absolutely due to the $tW$ loop threshold at $m_{b'} \simeq 250$ GeV. 
However, $a_{CP}=47(41)\%$ for other two modes, 
$b^{\prime} \to s\,(\gamma,g)$, Fig.\ref{figACPbpX}a,
are more reliable as theoretical predictions and for measurements as well.
Namely, the theoretical fact is that $a_{CP}(b^{\prime} \to s\,(\gamma,g))$ receive maximal values
for $m_{t'}\simeq 350$ GeV, 
which is shifted away from the $tW$ loop threshold. 
From the experimental point of view the best decay mode, 
out of $b' \to s$ modes, is certainly $b^{\prime} \to s\,\gamma$, 
because of the presence of a clean signal from the high energy single photon in the final state.
However, the bad point is the fact that $BR(b^{\prime} \to s\,\gamma)$, at 
$m_{t'}= 350$ GeV, could be as small as $10^{-6}$, Fig.\ref{figBRbpX}a, 
which is at the edge of the observable region for the LHC.
On the other hand, the shift down from $m_{t'}= 350$ GeV  to $m_{t'}= 300$ GeV increases  
$BR(b^{\prime} \to s\,\gamma)$  more than one order of magnitude 
(see Fig.\ref{figBRbpX}a),
and only slightly decreases $a_{CP}(b^{\prime} \to s\gamma)$ from $47\%$ to $42\%$ (Fig.\ref{figACPbpX}a). 
%Although the $BR(b^{\prime} \to s\,\gamma) \sim 10^{-6}$ is rather small, due to the fact that the high integrated 
%luminosity of $O( {\rm few}\, 100)\, {\rm fb^{-1}}$ could be achieved after several 
%years of LHC running \cite{CMS}, there is a possibility that the $b^{\prime} \to s\,\gamma$
%mode shall be observed and its $a_{CP}$ measured. 
Therefore, for $m_{b'}=300\,, 350$ GeV 
the required number of $b'$ quarks produced, in order to
obtain a $3\sigma$ CP violation effect, is $N_{b'}= 2.6 \cdot 10^6$, $4.1 \cdot 10^7$, 
respectively, which is a goal attainable after a few years of
operating the LHC \cite{CMS} at ${\cal O}({\rm few }\, 100)\,{\rm fb^{-1}}$. 

Comparing our estimate for $a_{CP}(b^{\prime} \to s\gamma)=47\%$
(see Fig.\ref{figACPbpX}a) 
with very recent predictions of ref. \cite{Arhrib:2009ew} we have found agreement up to
expected differences coming from the fitting procedure 
and the fitted CKM4 elements (\ref{300}-\ref{700}).

Discussing implications for the collider experiments we conclude: 
There are fair chances for the 4th generation quarks $b'$ and $t'$
to be observed at LHC and that their branching ratios could be measured. 
If LHC or future colliders discover 4th generation 
quarks at energies we assumed, it is highly 
probable that with well executed tagging large CP partial rate asymmetry could be found too.

\subsection*{Acknowledgments}
We would like to acknowledge M. Vysotsky for many stimulating comments, discussions
and careful reading of the manuscript. Comments from A. Soni are appreciated too.
B.M. acknowledges the hospitality of Department of Physics, Technion, Haifa and
G.E. wishes to thank the members of the Theoretical Physics Department at 
the Rudjer Bo\v{s}kovi\'{c} Institute, Zagreb,  for their kind hospitality. 
The work of B.M. and J.T. are supported by 
the Croatian Ministry of Science Education and Sports projects No. 
098-0982930-2864 and 098-0982930-2900, respectively. 
Work of G.E. received financial support from Technion, Haifa.
The work of J.~T. is also 
in part supported by the EU (HEPTOOLS) project under contract MRTN-CT-2006-035505.

%%%%%%%%%%%%%%%%%%%%%%%%%%%%%%%%%%%%%%%%%%%%%%%%%%%%%%%%%%%%%

%%%%%%%%%%%%%%%%%%%%%%%%%%%%%%%%%%%%%%%%%%%%%%%%%%%%%%%%%%%%%

\end{document}